\documentclass[12pt,preprint]{aastex}
\shorttitle{Scintillations and L\'evy flights}
\shortauthors{Boldyrev \& Gwinn}

\begin{document}
\input psfig.sty

\title{Scintillations and L\'evy flights through the
interstellar medium}

\author{Stanislav Boldyrev}
\affil{ Institute for Theoretical Physics, 
Santa Barbara, California 93106,\\ {\sf boldyrev@itp.ucsb.edu} }
\author{Carl Gwinn}
\affil{Department of Physics, University of California, Santa Barbara, 
California 93106, {\sf cgwinn@physics.ucsb.edu} }

%\date{\today}

\begin{abstract}
\vskip5mm
Temporal broadening of pulsar signals results from electron density
fluctuations in the interstellar medium that cause the radiation to
travel along paths of different lengths. The Gaussian theory of
fluctuations predicts that the  pulse temporal broadening should scale
with the wavelength as $\lambda^4$, and with the dispersion measure
(corresponding to distance to the pulsar) as $DM^2$. For large
dispersion measure, $DM>20 \,\mbox{pc/cm}^3$, the observed scaling is
$\lambda^4 DM^4$, contradicting the conventional theory. Although the
problem has existed for 30 years, there has been no resolution to this
paradox.

We suggest that scintillations for distant pulsars are caused by
non-Gaussian, spatially intermittent density
fluctuations with a power-like probability distribution. 
This probability distribution does
not have a second moment in a large range of density fluctuations,
and therefore the previously applied conventional Fokker-Planck theory
does not hold. Instead, we propose to apply the theory of L\'evy
distributions (so-called L\'evy flights). Using the scaling 
analysis (confirmed by numerical simulations of ray
propagation) we show that the observed scaling is recovered for 
large~$DM$, if the density differences, $\delta N$, 
have L\'evy distribution decaying as~$|\delta N|^{-5/3}$.
\end{abstract}

\section{Introduction}
%{\bf 1.} {\em Introduction.} 

Intensity fluctuations of pulsars radiation are caused by
scattering of radio waves by electron
density inhomogeneities in the interstellar medium. These fluctuations
are a signature of turbulent, non-equilibrium motion in the ISM, and
as the phenomenon of turbulence itself, they have withstood full 
theoretical understanding for decades, see e.g. reviews by~\citet{Sutton} 
and~\citet{Rickett1,Rickett2}. 
Observationally, the presence of electron density fluctuations 
leads, among other effects, 
to temporal and angular broadening of the pulsar image. These two
effects are naturally related -- due to fluctuations of the 
refraction index, different rays from a pulsar travel along paths 
of different shapes and the stronger the deviation of the path from
the straight line, the broader the pulsar image and the larger 
the time broadening of the arriving signal. Denoting the angular width 
of the image as~$\Delta \theta$, and using simple geometrical consideration, 
one estimates the arrival time broadening 
as~$\tau_d\approx (\Delta \theta)^2 d/c$, where~$d$ is the distance to the pulsar,
and $c$ is the speed of light, see more detailed discussion 
in~\citep{Blanford,Gwinn}. 
% see Fig.~\ref{fig1}.

A ray propagating through the interstellar medium encounters 
many
randomly distributed small ``prisms" on its way, that make the
scattering angle wander randomly.
%, and the distribution of the 
%arriving signal is a result of averaging over many such paths.
At each scattering event, the angle deflection is proportional 
to~$\lambda^2$ [see below], where $\lambda$ is the wavelength 
of the scattered
radiation. Taking into account that the scattering angle is 
small and exhibits the standard Gaussian random walk, we
estimate~$(\Delta \theta)^2\sim \lambda^4 d$, and the time delay scales 
as~$\tau_d \sim \lambda^4 d^2$, where $d$ plays the role of  
time in this random walk. The distance to the pulsar is
proportional to the dispersion measure, $DM$, and therefore this
relation can be checked experimentally. As has been consistently 
noted for more than 30 years,  observed scaling of
scintillations of distant pulsars, $DM>20\,\mbox{pc/cm}^3$, is far
from this simple theoretical prediction, instead, it is well described
by~$\tau_d\sim \lambda^4\, DM^4$, see e.g.,~\citep{Sutton,Rickett1}. 
Sutton proposed that scaling for longer lines of sight arose
from dramatically increased probability of intersection with 
strongly scattering HII regions.  In this sense,
he proposed that rare, large events dominated the 
line-of-sight averages.

The problem of scintillations was addressed by many authors who developed
thorough analytical models, see the discussion 
in~\citep{Tatarskii,Rumsey,Shishov,Lee1,Lee2,Goodman,Blanford,Lithwick}. These models 
account for both smooth and non-smooth density fluctuations, the latter can 
arise from turbulent cascades. 
The main object of the theories is the so-called projected 
correlator of density fluctuations. Denote as~$N({\bf r},t)$ the 
electron density,  
and~$\tilde N({\bf x},t)=\int^d
\mbox{d}z\, N({\bf r},t)$ its projection perpendicular to the 
distance~$d$. Here~${\bf
x}$ is a two dimensional vector in the plane perpendicular to the line
of sight, and $z$ is a coordinate along the line of sight, i.e.
${\bf r}=({\bf x}, z)$. Note that these theories all assume that the 
distribution of projected density fluctuations is Gaussian. 
The density and projected density correlators are related as:
\begin{eqnarray}
\langle \tilde N({\bf x}_1) \tilde N({\bf x}_2) \rangle
=\int\limits^d_0\int\limits^d_0 
\mbox{d}z\, \mbox{d}z'\,\langle N({\bf r}_1) N({\bf r}_2) 
\rangle ,
\end{eqnarray}
where both fields inside the brackets are taken at the same time.
Due to space homogeneity, these correlators depend only on the
difference of the coordinates, e.g., $\langle N({\bf r}_1) N({\bf r}_2) 
\rangle=\kappa({\bf r}_1-{\bf r}_2)$. Assuming that the density
fluctuations have finite correlation length~$l$, i.e. the $\kappa$
function decays fast for~$|{\bf r}_1-{\bf r}_2| >l$, we obtain
\begin{eqnarray}
\langle \tilde N({\bf x}_1) \tilde N({\bf x}_2) \rangle =
d \int\limits^\infty_0 \mbox{d}z\, \kappa({\bf x}_1-{\bf x}_2, z)\equiv d\, 
{\tilde \kappa}({\bf x}_1-{\bf x}_2).
\end{eqnarray}
It is easy to show that if in the inertial range of turbulent
fluctuations, $|r|\ll l$, the $\kappa$ function behaves as~$\kappa({\bf r})\approx
N_0^2[1-B(r/l)^\alpha]$, then the projected function is expanded 
as~${\tilde
\kappa(x,t)}\approx {\tilde N}_0^2[1-{\tilde B} (x/l)^{1+\alpha}]$. As an
estimate, one has~$ \langle N^2 \rangle=N_0^2 \sim {\tilde N}_0^2/l$, 
and $B$
and $\tilde B$ are of the order~1. The
analytical case corresponds to~$\alpha=1$, and in this 
case $\tau_d\sim \lambda^4 d^2$. In a general case,
the density field should not be analytic, and~$\alpha \neq 1$. For example, 
Kolmogorov turbulence would imply~$\alpha=2/3$.
Such different possibilities have been exhaustively analyzed 
in the literature, see e.g.~\citep{Lee1,Lee2,Goodman,Lambert}.
Rigorous consideration shows that in the non-analytic case, the
scaling of the broadening time changes. For~$\alpha\leq 1$, one 
obtains~
\begin{eqnarray}
\tau_d\sim \lambda^{2(\alpha +3)/(\alpha +1)} d^{(\alpha
+3)/(\alpha+1)},
\label{alpha<1}
\end{eqnarray} 
while for a more exotic case,~$\alpha>1$, 
one gets  
\begin{eqnarray}
\tau_d\sim \lambda^{8/(3-\alpha)}
d^{(3+\alpha)/(3-\alpha)}. 
\label{alpha>1}
\end{eqnarray}
In section \ref{ray-tracing} we present a simple derivation of 
these results. Since most observational data indicate
that~$\lambda$-scaling is close to~$\lambda^4$, neither possibility
 provides 
enough freedom for changing the $d$-scaling from $d^2$  to~$d^4$.

In the present paper we propose a new model, that fully exploits the
turbulent origin of the density fluctuations. We assume that the
statistics of the density fluctuations is not Gaussian, but 
highly intermittent, and that the probability density function (PDF) 
of density
differences has 
power-law decay, $P(\delta N)\sim |\delta N|^{-1-\beta}$. 
If this power-law 
distribution does not have a second moment ($\beta<2$), 
the Gaussian random walk approach does not work. Instead, we suggest 
to
use the theory of L\'evy distributions, see~\citep{Shlesinger}. 
Physically, the possibility
of power-law density distribution seems rather natural for strong
turbulent fluctuations. Indeed,  the ISM turbulence can be  
near-sonic, i.e. velocity and density fields can develop shock
discontinuities. From the theory of shock turbulence (Burgers
turbulence) one knows that shocks or large negative velocity
gradients have a power-law distribution,~\citep{Polyakov,E,Boldyrev}. 
Jump conditions on a shock
then show that the velocity and density discontinuities are 
proportional to
each other, therefore density jumps may also have 
power-law distribution. 
%As another possibility, rather intermittent 
%electron density fluctuations can originate in  highly 
%ionized H\,II regions.
Taking the L\'evy distribution of the density fluctuations 
as a working conjecture, 
we demonstrate that the scaling of the
broadening time with respect to~$d$ is sensitive to the 
exponent of the distribution, $\beta$, and the scaling~$\tau_d\sim 
\lambda^4 d^4$ is reproduced for~$\beta=2/3$.

In the next section we review the ray-tracing model of pulse
propagation, considered before by~\citet{Williamson1,Williamson2,Blanford}. 
In particular, we re-derive the 
results cited above 
for the Gaussian density fluctuations in a general, non-analytic case. 
In Section~\ref{levy} we 
%consider only smooth density fields, 
%but 
apply the model to the non-Gaussian, 
L\'evy distributed density fluctuations. We then
numerically calculate the distribution of pulse-arrival times
in the case of a smooth density field,  
and demonstrate that if~$P(\delta N)\sim |\delta N|^{-5/3}$,  
the width of this distribution changes with the
distance to the pulsar as~$\lambda^4 d^4$, in agreement with our scaling arguments. 
Conclusions and future research are outlined
in Section~\ref{conclusions}.

\section{Ray-tracing method}
\label{ray-tracing}
This method is applicable in the limit of geometrical optics, i.e.
when the wave length is much smaller than the characteristic size of
density inhomogeneities~\citep{Landau}. This rather effective method 
was applied to the 
problem of 
scintillations by~\citet{Williamson1,Williamson2,Blanford}; 
we present it here in the form that allows us to apply it 
in the next section to L\'evy walks.
In the limit considered, signal propagation can be characterized by 
rays,  
${\bf r}(t)$, along which wave packets travel similar 
to  particles obeying the following system of Hamilton equations:
\begin{eqnarray}  
\dot {\bf r}&=&\partial \omega(k,r)/\partial {\bf
k},\nonumber \\
\dot {\bf k}&=&-\partial \omega(k,r)/\partial {\bf
r}.
\label{rays}
\end{eqnarray}
In this representation, $\omega$ plays the role of Hamiltonian,  
$\omega^2=\omega^2_{pe}(r)+k^2c^2$, where~$\omega^2_{pe}(r)=4\pi
N(r) e^2/m_e$ is the electron plasma frequency and $\bf k$ 
is a wave vector. Differentiating the first equation in~(\ref{rays})
with respect to~$t$ and using the second equation one obtains:
\begin{eqnarray}
\ddot {\bf r}=-2\pi c^2 \lambda^2 r_0 \partial N(r)/\partial {\bf r},
\label{ray-eq}
\end{eqnarray}
where~$r_0=e^2/m_e c^2$ is the classical radius of electron. Taking
into account that the ray propagates at small angles to the line of
sight, chosen as the $z$-axis, we are interested in ray displacement
in the perpendicular, $\bf x$ direction, and instead of time
we will use $z$ variable,~$z=ct$. Consider now two rays, separated by
a vector~$\Delta {\bf x}$ in the direction perpendicular to 
the $z$-axis. As follows from~(\ref{ray-eq}), this vector
obeys the following equation
\begin{eqnarray} 
\frac{d (\Delta {\bf x})}{d z}&=&\Delta {\bf v}, \nonumber \\
\frac{d (\Delta {\bf v})}{d z}&=&A\,\Delta \frac{\partial N(x,z)}{\partial 
{\bf
x}}, 
\label{rays2}
\end{eqnarray}
where~$A=-2\pi \lambda^2 r_0$, and~$\Delta {\bf v}$ is an auhiliary  
variable
having the meaning of velocity of beam spreading in the ${\bf x}$
direction, clearly~$\Delta \theta \sim |\Delta {\bf v}|$. 
%This is 
%also a Hamiltonian system of
%equations but with time-dependent ($z$-dependent) Hamiltonian. 
Let us
now assume that the electron density is a Gaussian random function with
the correlation length~$l$. Then, $\Delta {\bf v}(z)$ is a Gaussian
random walk, whose elementary time step has the length~$l$. Since 
we are interested in very large propagation distances, $z \gg l$, and
the scattering angles are very small, one
can effectively assume that the random density is short-time
correlated, i.e., the characteristic ``$z$-time'' 
of change of vectors~$\Delta
{\bf v}$ and ${\Delta {\bf x}}$ is much larger than~$l$. 

The diffusion coefficient for this random walk is
\begin{eqnarray}
D=-A^2\frac{d^2 \tilde \kappa(\Delta x)}{d (\Delta x)^2}
\sim  \lambda^4 r_0^2 N_0^2 \left({\Delta x\over l}\right)^{\alpha-1} 
{1\over l},
\label{D}
\end{eqnarray}
and the diffusion is described by $(\Delta \theta)^2\sim D z$.
We however observe that the diffusion coefficient depends of the
distance~$\Delta x$, and its behavior differs qualitatively for
$\alpha<1$ and $\alpha>1$. In the first case, $\alpha<1$, diffusion 
is larger for
smaller distances, therefore two rays are effectively attracting each
other in the course of propagation. This means that at some point 
the geometrical ray picture will break down, and one needs to 
consider the effects of interference (interaction) of different rays. 
This happens when the beam is compressed 
to the size limited by 
the uncertainty condition in the perpendicular 
direction , $k\, \Delta \theta \Delta x\sim 1$. 
Upon substituting $\Delta \theta \sim D^{1/2} z^{1/2}$, and using 
the expression 
for the diffusion coefficient~(\ref{D}), 
we can obtain the minimal size of contraction, and, 
equivalently, the diffraction angle 
corresponding to the aperture of this size. Assuming that the
contraction happens at about half the distance  
between the pulsar and the Earth, $z\sim d/2$, we find:
\begin{eqnarray}
(\Delta \theta)^2 \sim \left[N_0^4 r_0^4 l^{-2\alpha}
\lambda^{2(\alpha+3)} d^{2}\right]^{1/(\alpha+1)}, \quad \alpha<1. 
\label{theta1}
\end{eqnarray}
Recalling now that~$\tau_d\sim (\Delta \theta)^2 d$, 
we recover the
result~(\ref{alpha<1}). In the second case, $\alpha>1$, the rays
effectively repel, so geometrical optics does not break down. In
this case $\Delta x \sim \Delta v \, z \sim D^{1/2} z^{3/2}$. This
equation gives 
\begin{eqnarray}
(\Delta \theta)^2 \sim \left[ N_0^4 r_0^4 l^{-2\alpha}
\lambda^8 d^{2\alpha}\right]^{1/(3-\alpha)}, \quad 1\leq \alpha <3,
\label{theta2}
\end{eqnarray}
which agrees with the result~(\ref{alpha>1}). Both expressions give 
the same result for the analytic case, $\alpha=1$. The above standard 
results have been obtained by many authors and by a 
variety of different methods, 
see e.g.,~\citep{Williamson1,Lee1,Lee2,Goodman,Blanford}. 
As we mentioned in
the introduction, neither one of the expressions~(\ref{theta1}) 
or~(\ref{theta2})
allows us to recover the observed scaling $\tau_d\sim \lambda^4 d^4$. 
In
the next section we address the problem, assuming that the 
density-difference
distribution has a slowly decaying power-law tail, such that the
second moment of the distribution does not exist. In this case the
diffusion approximation does not hold, and one needs to work directly
with Eq.~(\ref{rays2}) to establish the scaling of the probability of
pulse arrival times.

\section{L\'evy model for scintillations}
\label{levy}
In previous section we implicitly used the central limit theorem, 
which states that the sum of many independent random variables has 
Gaussian distribution if second moments of these variables exist. 
More precisely, a convolution of many 
distribution functions that have second moments, converges to 
an appropriately rescaled Gaussian distribution. Therefore, the 
convolution of two Gaussian functions is a Gaussian function again. 
One
can generalize this question for distribution functions without
finite second moments: if their convolution converges, what 
is going to be
the limit? The answer is the so-called L\'evy 
distribution~\citep{Shlesinger}. 
As is the Gaussian distribution, the L\'evy distribution is stable: 
convolution of this distribution with itself 
gives the same distribution after proper rescaling.
In other words, if two independent random variables are drawn from 
a L\'evy distribution, their
 sum has the same distribution, appropriately rescaled. Analogously 
to a 
Gaussian random walk, 
a sum of independent, L\'evy distributed random variables is 
called a L\'evy walk or L\'evy flight. The latter name reflects 
the highly intermittent behavior of a typical L\'evy trajectory:
it has sudden large jumps or ``flights,'' 
see Fig.~\ref{flights}. L\'evy flights 
are common in completely different random systems and often 
replace 
diffusion in turbulent systems. For example, a particle 
exhibiting a Brownian
random motion in an equilibrium fluid, exhibits a L\'evy walk in a
turbulent fluid. For a variety of further illustrations 
see~\citep{Shlesinger}.

If a random variable~$y$ has 
a L\'evy probability density,~$P(y)$, then the
Fourier transform of this distribution (the characteristic function)
has the form:
\begin{eqnarray}
\Phi(\mu) = \int\limits^{\infty}_{-\infty} \mbox{d}y\, P(y)\exp(i\mu
y)=\exp(-C|\mu|^{\beta}), 
\label{phi}
\end{eqnarray}
where~$0<\beta<2$, and $C$ is some positive constant. 
For $\beta=2$ we recover a Gaussian distribution. 
This
formula can be taken as the definition of a symmetric L\'evy walk. One
can verify that~$P(y)\sim |y|^{-1-\beta}$ as~$|y|\to \infty$. Of
course, a
distribution of a physical quantity usually has a second moment. 
This does not contradict our case, since the far tails of the PDF,
which are not described by the L\'evy formula, make the dominant
contribution to the second moment. However, if we are
interested in effects caused by small fluctuations, $y\ll y_{rms}$, 
it is the central part of the PDF that is important.

The characteristic function of a convolution of $n$ L\'evy 
distributions is just a product of $n$ characteristic
functions~(\ref{phi}). We therefore conclude that the sum of~$n$
L\'evy distributed random variables has the distribution
\begin{eqnarray}
P_n(y)=P(yn^{-1/\beta})n^{-1/\beta}.
\label{P}
\end{eqnarray}
This is the demonstration of the convolution {\em stability} 
of the L\'evy distribution. Formula~(\ref{P}) teaches us that the
displacement~$y$ of the L\'evy random walk scales with the number of
steps as~$y\sim n^{1/\beta}$. In the Gaussian case, $\beta=2$, 
we recover the well known diffusion result.

We now would like to apply this result to our scintillation problem.
Let us {\em assume} that the dimensionless density
difference~$\Delta N(x)/N_0$ 
has a L\'evy distribution with parameter~$\beta$. 
We then obtain from~(\ref{rays}), $\Delta {\bf v}\sim -A\Delta N$, 
and [compare this result to~(\ref{D})!]: 
\begin{eqnarray}
(\Delta \theta)^2\sim  \lambda^4 r_0^2 N_0^2 
\left({\Delta x\over l}\right)^{\alpha-1}
\left({z\over l}\right)^{2/\beta}.
\label{levy-diffusion}
\end{eqnarray}
In this formula, $\alpha$ describes the scaling of the 
density fluctuations with
distance, while $\beta$ is the exponent of the power-law decay of the
density-difference probability distribution function. 
The scaling in Eq.~(\ref{levy-diffusion}) is understood not in 
the sense of averaging (the moments
of $\Delta \theta$  of the order higher than $\beta$ do not satisfy
this scaling), but in the sense of scaling of the central part of the 
distribution~$P_z(\Delta \theta)$. We now proceed exactly as we did 
in the
derivation of formulae (\ref{theta1}) and (\ref{theta2}), and 
obtain for~$\alpha<1$: 
\begin{eqnarray}
(\Delta \theta)^2 \sim \left[N_0^4 r_0^4 l^{2-2\alpha -4/\beta} 
\lambda^{2(\alpha+3)} d^{4/\beta}\right]^{1/(\alpha+1)},  
\label{levy1}
\end{eqnarray}
and for~$1\leq \alpha <3$:
\begin{eqnarray}
(\Delta \theta)^2 \sim \left[ N_0^4 r_0^4 l^{2-2\alpha-4/\beta}
\lambda^8 d^{2\alpha-2+4/\beta}\right]^{1/(3-\alpha)}.
\label{levy2}
\end{eqnarray}
In the smooth (analytic) case, $\alpha=1$, the scaling of the 
time broadening is
\begin{eqnarray}
\tau_d \sim \left(N_0^2 r_0^2 l^{-2/\beta}/c\right)
\lambda^4 d^{(2+\beta)/\beta}. 
\label{levy-smooth}
\end{eqnarray}
We see
that this scaling 
is sensitive to the exponent of the power distribution of the
density fluctuations. This result was obtained by rather general
arguments, and describes the scaling of the arrival time
 distribution, rather than the moments of this
distribution. Observations measure precisely the time width of 
the arriving signal, not its moments, i.e. they infer
exactly the quantity corresponding to scaling~(\ref{levy-smooth}).  

In the rest of this section we would like to verify the
scaling~(\ref{levy-smooth}), by numerical simulation of 
Eq.~(\ref{rays2}). Our simulations 
also provide the time-shape of the arriving signal. 
Let us assume that the distance to the pulsar, $d$, is much larger
than the scale of an elementary scatter, $l$, i.e., $n=d/l \gg 1$, where 
$n$~is the number of scattering events. At each
scattering event, the angle of the ray changes  
by~$\delta \theta\ll 1$, where $\delta \theta$ is a L\'evy-distributed 
random variable. [We denote $\delta N$ and $\delta \theta$ 
the characteristic changes of the density field and of the angle of 
propagation on one scattering segment of length~$l$ along the 
line of sight. This should not be confused with the changes of these 
variables between two {\em different} rays in a 
perpendicular plane~${\bf x}$, denoted by $\Delta$'s.] 
The time delay (compared to the
straight propagation) introduced by each scattering segment 
is~$\delta \tau_d \sim l\theta^2/c$. We need to find the
probability distribution of the total travel time delay:
\begin{eqnarray}
\tau_d=\frac{l}{c} \sum\limits_{m=1}^{n}
\theta_m^2=\frac{l}{c} \sum\limits_{m=1}^{n}
\left[\sum\limits_{s=1}^{m} \delta \theta_s \right]^2,
\label{tau}
\end{eqnarray}
assuming that each $\delta \theta_s$ 
[where $\delta \theta_s \approx -A\delta N$ due to
(\ref{rays2})] is distributed 
identically,
independently, and according to the L\'evy law~(\ref{phi}) 
with~$\beta=2/3$. This random variable can be generated in the
following manner, see~\citep{Klafter}. 
Choose two positive numbers~$a>b>1$. 
Let $\delta  \theta=\theta_0 b^i$ with probability~$P(i)=(a-1)/(2a^{i+1})$, 
and $\delta \theta=-\theta_0 b^i$ with 
the same probability, where $i=0,1,2\dots$. This is the so-called 
Weierstrass self-similar random walk, that can be considered as a discrete analog of 
the L\'evy walk with $\beta=\ln(a)/\ln(b)$.

In Fig.~\ref{pdf} we plot intensity of the arriving 
signal (the number of arriving rays) as a function of time.  
The arrival time was calculated with the aid of~(\ref{tau}), where in 
the distribution of $\delta \theta$ we have chosen $\beta=2/3$, $b=4$, and
$\theta_0=0.0002$.  
We considered the number of scattering events (the distance to the pulsar) 
to be~$n_1=100$, $n_2=100\times 2^{1/4}\approx 119$, and 
$n_3=100\times 2^{-1/4} \approx 84$. 
%The $\delta \theta$ amplitude
% was chosen $\theta_0=0.0002$. 
From Fig.~\ref{pdf} one can see that the widths of the curves 
(estimated at the half of their maximum values) indeed differ by 
a factor of 2, as the scaling $\tau_d\sim \lambda^4 d^4$ would predict for 
these distances.

\section{Conclusions}
\label{conclusions}
In conclusion, we suggest a novel explanation for the observed scaling 
of time broadening of pulsar signals for large distances 
(large dispersion measures, $DM>20\,\mbox{pc/cm}^3$), 
$\tau_d \sim \lambda^4 d^4$. The central concept is that 
the density fluctuations in the interstellar medium have 
a L\'evy probability distribution function that has power-law 
decay and 
does not have a second moment. The angle of pulse propagation, 
deviated by these density 
fluctuations, exhibits not a conventional Brownian motion, 
but rather a L\'evy flight. The exponent 
$\beta$ is the parameter of the  
probability distribution of density differences, and 
the pulse broadening time 
is rather sensitive to it, as is described by our main 
formulae~(\ref{levy1}) and (\ref{levy2}).
The scaling $\tau_d \sim \lambda^4 d^4$ is 
recovered for $\beta=2/3$, i.e. for the~$|\delta N|^{-5/3}$ decay of 
the distribution function of density differences. 
This tail of the PDF appears as a result of turbulent density 
fragmentation, and it would be highly desirable to develop an 
analytical explanation 
for it. This is  
a concrete prediction of our model for the turbulence in 
the ISM, that can in principle be checked numerically.

%{
%%\columnwidth=3.2in
%\begin{figure} [tbp]
%\centerline{\psfig{file=ray.ps,width=7in,angle=-90}}
%\vskip5mm
%\caption{A typical L\'evy random walk trajectory exhibits sudden large 
%deviations, ``flights.'' In the case of ray propagation through 
%the ISM, the ray angle commits a L\'evy flight. Strong angle 
%deviations 
%occur when the ray encounters regions of large electron density 
%inhomogeneities, such as shocks or HII regions. (Angular scale 
%is arbitrary.)
%} 
%\label{flights}
%\end{figure}
%}

%{
%%\columnwidth=3.2in
%\begin{figure} [tbp]
%\centerline{\psfig{file=pdfs.ps,width=7in,angle=-90}}
%\vskip5mm
%\caption{Numerical calculation of the number of arriving rays vs 
%time (time units are arbitrary). 
%We used  
%formula~(\ref{tau}), and the L\'evy distributed density 
%fluctuations with $\beta=2/3$. 
%We calculated arrival times of $10^6$ rays for three 
%different distances to the source, $n_1=100$, $n_2=84\approx 
%100\times 2^{-1/4} $, and $n_3=119\approx 100\times 2^{1/4}$.
%One observes that the width of the plot ``n=84'' is twice as small, and 
%the with of the plot ``n=119'' is twice as large, as the width of 
%the plot ``n=100.'' This corresponds to the 
%scaling $\tau_d\sim d^4$. 
%} 
%\label{pdf}
%\end{figure}
%}

%\end{multicols}

\newpage

{
\begin{figure} [tbp]
\centerline{\psfig{file=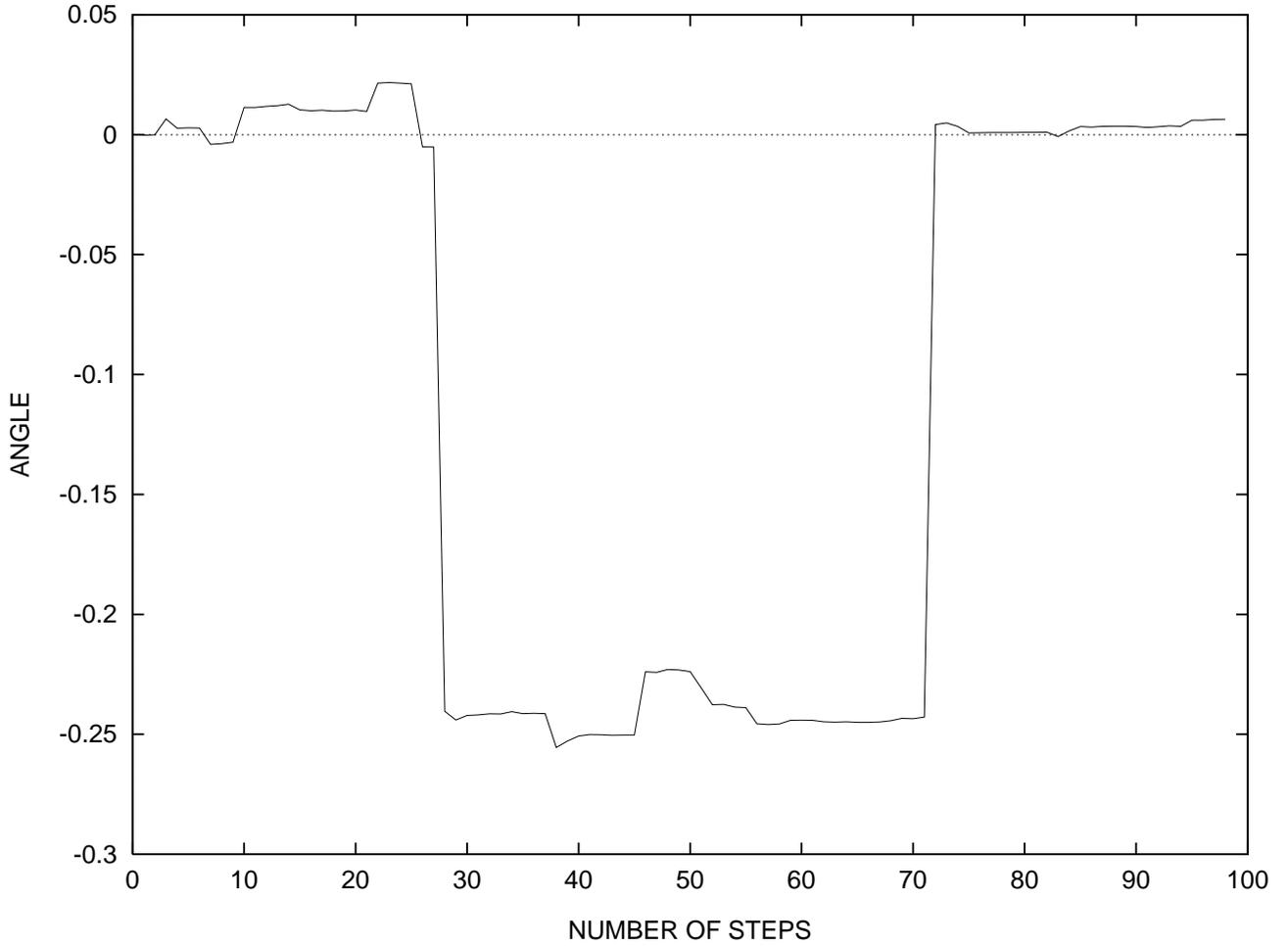,width=7in,angle=-90}}
\vskip5mm
\caption{A typical realization of a L\'evy random walk. 
The trajectory exhibits sudden 
large 
deviations, ``flights.'' In the case of ray propagation through 
the ISM, the ray angle performs a L\'evy walk. Large angular  
deviations occur when the ray encounters regions of large electron density 
inhomogeneities, such as shocks or HII regions. (Angular scale 
is arbitrary.)
} 
\label{flights}
\end{figure}
}

{
\begin{figure} [tbp]
\centerline{\psfig{file=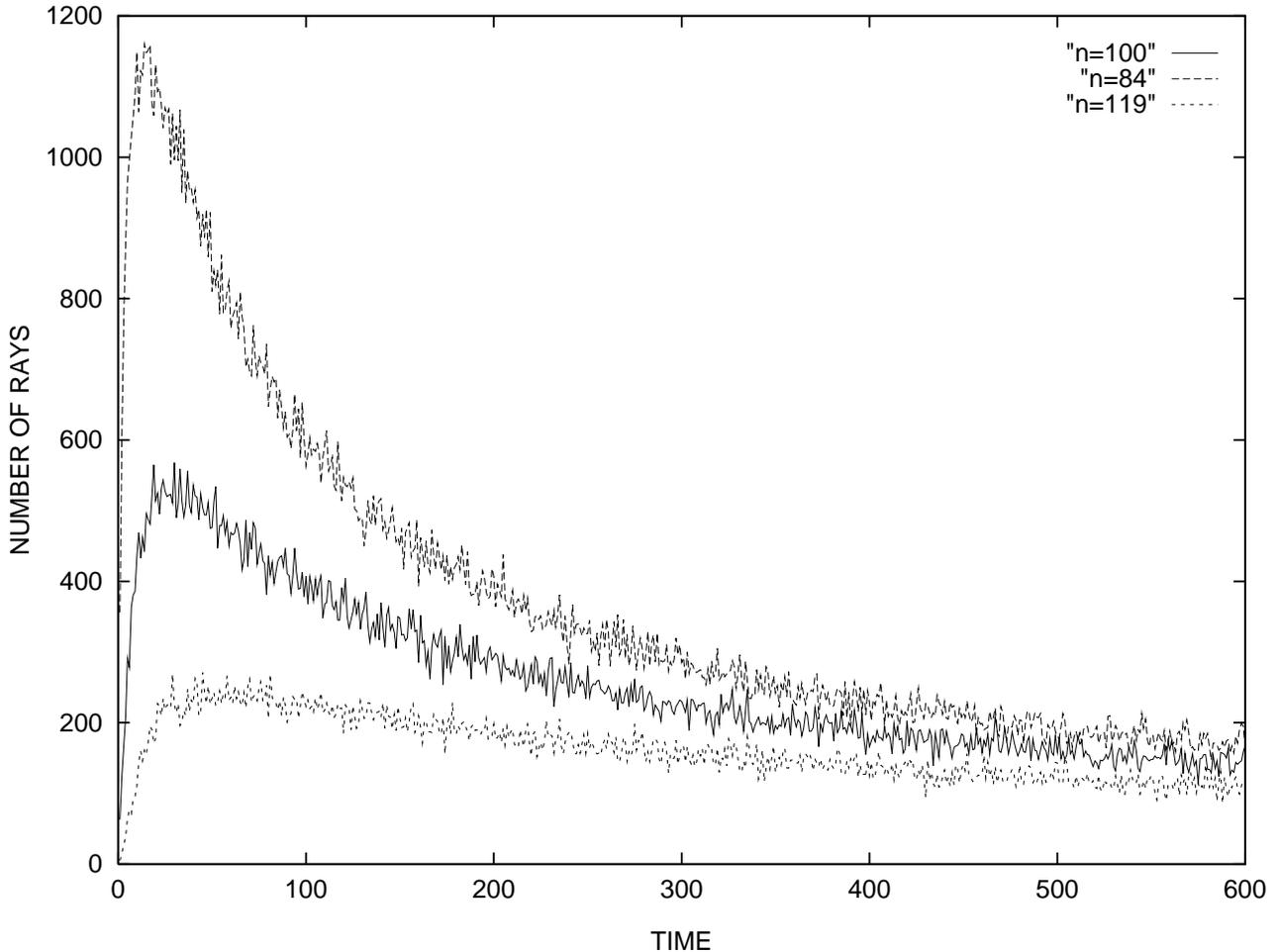,width=7in,angle=-90}}
\vskip5mm
\caption{Numerical calculation of the number of arriving rays vs 
time (time units are arbitrary). 
We used  
formula~(\ref{tau}), and the L\'evy distributed density 
fluctuations with $\beta=2/3$. 
We calculated arrival times of $10^6$ rays for three 
different distances to the source, $n_1=100$, $n_2=84\approx 
100\times 2^{-1/4} $, and $n_3=119\approx 100\times 2^{1/4}$.
One observes that the width of the plot ``n=84'' is twice as small, and 
the width of the plot ``n=119'' is twice as large, as the width of 
the plot ``n=100.'' This corresponds to the 
scaling $\tau_d\sim d^4$. 
} 
\label{pdf}
\end{figure}
}


\begin{thebibliography}{99}

\bibitem[Blanford \& Narayan(1985)]{Blanford} Blanford, R. and Narayan, R.  
   1985, MNRAS, {\bf 213}, 591.
\bibitem[Boldyrev(1998)]{Boldyrev} Boldyrev, S. 1998, 
   Phys. Plasmas, {\bf 5}, 1681.
\bibitem[Gochelashvily \& Shishov(1975)]{Shishov} Gochelashvily, K. S., and 
   Shishov, V. I. 1975, Opt. Quant. Electron., {\bf 7}, 524.
\bibitem[Goodman \& Narayan(1985)]{Goodman} Goodman, J. and Narayan, R.  
   1985, MNRAS, {\bf 214}, 519.
\bibitem[E~{\em et al}(1997)]{E} E, W., Khanin, K., Mazel, A., 
   and Sinai Ya. 1997, Phys. Rev. Lett., {\bf 78}, 1904.
\bibitem[Gwinn, Bartel, \& Cordes(1993)]{Gwinn} Gwinn, C.~R., Bartel, N., 
   and Cordes, J.~M. 1993, \apj, {\bf 410}, 673.
\bibitem[Klafter, Zumofen, \& Shlesinger(1995)]{Klafter} Klafter, J., Zumofen, G., 
   and Shlesinger, M. S., (1995), in {\em L\'evy flights and related topics in
   physics}, (Springer, Berlin). 
\bibitem[Lambert \& Rickett(2000)]{Lambert} Lambert, H.~C. and  
   Rickett, B.~J. 2000, \apj, {\bf 531}, 883.
\bibitem[Lifshitz, Landau, \& Pitaevsky(1995)]{Landau} Lifshitz, E. M., 
   Landau, L.~D., and Pitaevsky L. P., (1995) {\em Electrodynamics 
   of Continuous Media}, vol. VIII,
   (Butterworth-Heinemann).
\bibitem[Lee \& Jokipii(1975a)]{Lee1} Lee, L.~C. and Jokipii, J.~R. 1975(a), 
   \apj,~{\bf 196}, 695.
\bibitem[Lee \& Jokipii(1975b)]{Lee2} Lee, L.~C. and Jokipii, J.~R. 1975(b), 
   \apj,~{\bf 201}, 532.
\bibitem[Lithwick \& Goldreich(2001)]{Lithwick} Lithwick, Y. and 
   Goldreich, P. 2001, \apj~{}; astro-ph/0106425.
\bibitem[Polyakov(1995)]{Polyakov} Polyakov, A.~M. 1995, Phys. Rev. 
   E, {\bf 52}, 6183.
\bibitem[Rickett(1977)]{Rickett1} Rickett, B.~J. 1977, Ann. Rev. 
   Astron. Astrophys., {\bf 15}, 479.
\bibitem[Rickett(1990)]{Rickett2} Rickett, B.~J. 1990, Ann. Rev. 
   Astron. Astrophys., {\bf 28}, 561.
\bibitem[Rumsey(1975)]{Rumsey} Rumsey, V. H. 1975, Radio Sci., {\bf 10}, 107. 
\bibitem[Shlesinger, Zaslavsky, \& Frisch(1995)]{Shlesinger} 
   Shlesinger, M.,  Zaslavsky, G., and Frisch, U., Eds., (1995), 
   {\em L\'evy flights and related topics in physics}, 
   (Springer, Berlin). 
\bibitem[Sutton(1971)]{Sutton} Sutton, J.~M. 1971, MNRAS, {\bf 155}, 51.
\bibitem[Tatarskii \& Zavorotnyi(1980)]{Tatarskii} Tatarskii, V. I. 
   and Zavorotnyi, V. U. 1980, Progress in Optics, p. 207, ed. Wolf, E., vol.
   XVIII. 
\bibitem[Williamson(1972)]{Williamson1} Williamson, I.~P. 1972, 
   MNRAS, {\bf 157}, 55.
\bibitem[Williamson(1973)]{Williamson2} Williamson, I.~P. 1973, 
   MNRAS, {\bf 163}, 345.


\end{thebibliography}
\end {document}